\documentstyle[12pt,epsfig]{elsart}

\begin{document}
\newcommand{\beq}{\begin{equation}}
\newcommand{\eeq}{\end{equation}}
\newcommand{\beqa}{\begin{eqnarray}}
\newcommand{\eeqa}{\end{eqnarray}}
\begin{frontmatter}
\title{Nuclear stopping and flow in heavy ion collisions and the in-medium NN cross section}
\author{T. Gaitanos$^1$, C. Fuchs$^2$, H.~H. Wolter$^3$}
\address{${}^{1}$Laboratori Nazionali del Sud INFN, I-95123 Catania, Italy \\ 
${}^{2}$Institut f\"ur Theoretische Physik der 
Universit\"at T\"ubingen, D-72076 T\"ubingen, Germany\\ 
${}^{3}$Dept. f\"ur Physik, Universit\"at M\"unchen, 
D-85748 Garching, Germany}  
\begin{abstract}
We present transport calculations for heavy ion reactions in which the mean field and
the in-medium nucleon-nucleon cross section are consistently based on the 
same effective interaction, i.e. the in-medium T-matrix from
microscopic Dirac-Brueckner calculations. Doing so, the stopping in central
reactions in terms of the recently proposed 
$var_{\rm tl}$ observable and the correlation to the behavior of the directed flow 
is investigated. The relation to the nuclear shear viscosity is discussed.
\end{abstract}
\end{frontmatter}
One of the primary goals in studying heavy ion reactions at intermediate
energies is the investigation of the nuclear equation-of-state (EOS) 
at supra-normal densities and/or high
temperatures \cite{fopi1}. In a hydrodynamical picture 
the time evolution of such a
reaction can be understood in terms of pressure gradients which build
up in the compressed zone and drive the dynamics. Originally it was 
therefore expected that a direct determination of the nuclear EOS via 
collective flow observables should be possible \cite{hydro1}. 
However, the situation 
turned out to be more complex: The system does not behaves like an ideal
fluid but binary collisions of the nucleons lead to a viscous
behavior. Moreover, over most of the reaction time, in particular
during the compressional phase, the system is even out of {\it local} 
equilibrium \cite{temp,gait01}. E.g. experimental evidence for 
incomplete stopping even in central reactions has been reported
in \cite{fopi3}. Thus the hydrodynamical limit is not reached in
relativistic heavy ion reactions except of the extreme ultra-relativistic
regime \cite{star}.

On the other hand, microscopic transport models turned out to be 
an adequate tool for the description of the reactions dynamics at
intermediate energies. The physical input of such Boltzmann type 
semi-classical models are the nuclear 
mean field $U$ and the nucleon-nucleon (NN) cross section
$\sigma$. Both are determined by the effective 
two-body interaction in the medium,
i.e. the in-medium T-matrix;  $U\sim \Re T\rho,~~
\sigma\sim \Im T$, respectively $d\sigma/d\Omega\sim | T|^2$. 
However, in most practical applications phenomenological mean fields 
and cross sections are used. Adjusting the known bulk properties
of nuclear matter around the saturation point one tries to 
constrain the models for supra-normal densities 
with the help of heavy ion reactions 
\cite{dani00,larionov00}. Medium modifications of the NN cross section 
are often disregarded which works, from a practical point of view, 
astonishingly well. However, in particular kinematics regimes a 
sensitivity of dynamical observables such as in-plane ($v_1$) and 
elliptic ($v_2$) flow \cite{gale} or transverse energy transfer 
\cite{dancross} to the NN cross section has been observed. 

In this letter we present calculations were both, the mean
field and the in-medium cross sections are consistently based on the
relativistic Dirac-Brueckner (DB) T-matrix. 
The standard Dirac-Brueckner ladder approximation \cite{DBT1} accounts 
for two body-correlations to lowest order in the Brueckner hole-line expansion. 
Microscopic DB mean fields 
have already been successfully applied to heavy ion reactions
\cite{gait01} but a consistent treatment of mean field and cross section 
has not been performed so far. In a consistent
truncation scheme two-body correlations are maintained at the 
same level in the drift term and the
collision integral of the kinetic equation \cite{btm}.

At finite densities the NN cross section is suppressed 
for two reasons: (1) Kinematics, i.e. the cross 
section scales with the in-medium nucleon mass as $\sigma_{\rm eff}
\sim (M^*/M)^2 \sigma$ \cite{malfliet88,fuchs01}. (2) At low energies 
an additional suppression due to Pauli-blocking of the {\it
intermediate} states in the Bethe-Salpeter (BS) equation 
\beq
T= V + iVGGQT
\label{BS}
\eeq
appears. The Pauli-Operator $Q=(1-f)(1-f)$ in (\ref{BS}) 
projects intermediate states onto unoccupied phase space
regions. The mean field, respectively the
Hartree-Fock self-energy, is obtained by a summation over the Fermi-sea 
$\Sigma = -i tr[T G_F ]$. This leads to dressed quasi-particles carrying 
effective masses and kinetic momenta
        \begin{equation}                                
        M^{\ast} = M + {\rm \Re}\Sigma_s(k_F) \, , \qquad 
        k^{\ast\mu} = k^{\mu} + {\rm \Re}\Sigma^{\mu}(k_F) 
\label{effmass}
        \end{equation}
given by the real part of the, in general complex, self-energy. 
For on-shell scattering the NN cross section is given by 
\beq
d\sigma = \frac{(M^*)^4}{s^* 4 \pi^2} | T |^2 
d\Omega~.
\label{sig1}
\eeq
In the present work we use both, mean fields 
and in-medium cross sections obtained in DB
calculations of the T\"ubingen group \cite{DBT1,fuchs01} 
utilizing the Bonn A one-boson-exchange potential as the 
bare NN interaction $V$  which yields very reasonable saturation
properties for infinite nuclear matter.
\begin{figure}[t]
\unitlength1cm
\begin{picture}(8.,7.3)
\put(3.0,0.3){\makebox{\epsfig{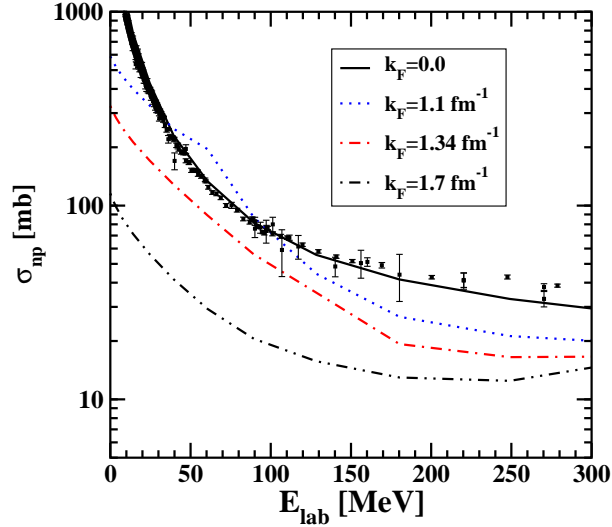}}}
\end{picture}
\caption{Elastic in-medium neutron-proton cross sections 
at various Fermi momenta 
$k_{F}$ as function of the laboratory energy $E_{lab}$. The free cross
section ($k_{F}=0$) is compared to the experimental total $np$ cross
section \protect\cite{pdg}.
}
\label{fig1}
\end{figure}
Fig. 1 shows the energy dependence of the in-medium 
neutron-proton $(np)$ cross section \cite{fuchs01} at 
Fermi momenta $k_F = 0.0, 1.1,1.34,1.7 fm^{-1}$, 
corresponding to $\rho \sim 0,0.5,1,2\rho_0$ ($\rho_0=0.16 fm^{-3}$ 
is the nuclear matter saturation density). 
The presence of the medium leads to a substantial 
suppression of the cross section which is most pronounced at 
low laboratory energy $E_{\rm lab}$ and high densities where, in addition to 
the $(M^*/M)^2$ scaling of (\ref{sig1}) 
the Pauli-blocking of intermediate states is 
most efficient. At larger $E_{\rm lab}$ asymptotic values of 
15-20 mb are reached. However, not only the total cross section 
but also the angular distributions are affected by the presence of the 
medium. The initially highly forward-backward 
peaked  $np$ cross sections becomes much more isotropic at finite densities 
\cite{fuchs01} which is mainly do to the Pauli suppression of 
soft modes ($\pi$-exchange) and corresponding higher partial waves in 
the T-matrix. 

Heavy ion collisions are described by the relativistic 
Boltzmann-Uehling-Uhlenbeck (RBUU) transport equation, 
which has the form \cite{btm}
\begin{eqnarray}
& & \left[ 
k^{*\mu} \partial_{\mu}^{x} + \left( k^{*}_{\nu} F^{\mu\nu} 
+ M^{*} \partial_{x}^{\mu} M^{*}  \right) 
\partial_{\mu}^{k^{*}} 
\right] f(x,k^{*}) = \frac{1}{2(2\pi)^9} \nonumber\\
& & \times \int \frac{d^3 k_{2}}{E^{*}_{{\bf k}_{2}}} 
             \frac{d^3 k_{3}}{E^{*}_{{\bf k}_{3}}}
             \frac{d^3 k_{4}}{E^{*}_{{\bf k}_{4}}} W(kk_2|k_3 k_4)   
 \left[ f_3 f_4 \tilde{f}\tilde{f}_2 -f f_2 \tilde{f}_3\tilde{f}_4 
\right]
\label{rbuu} 
\end{eqnarray}
where the single particle distribution function is given by 
$f_i = f({\bf x},t;k^{*}_i)$
and the hole-distribution by $\tilde{f}_i = (1-f({\bf x},t;k^{*}_i))$. The 
collision integral accounts explicitly for final state Pauli-blocking 
while the in-medium scattering amplitude includes the  Pauli-blocking 
of intermediate states. 

The dynamics of the lhs of eq.(\ref{rbuu}), the drift term, is 
determined by the mean field. Here the 
attractive scalar field $\Sigma_S$ enters via the effective mass $M^{*}$ 
and the repulsive vector field $\Sigma_\mu$ via 
kinetic momenta $k^{*}_{\mu}$ and via the field tensor 
$F^{\mu\nu} = \partial^\mu \Sigma^\nu -\partial^\nu \Sigma^\mu$. 
The in-medium cross sections enter into the collision integral 
via the transition amplitude  
\beq
W = (2\pi)^4 \delta^4 \left(k + k_{2} -k_{3} - k_{4} \right) 
(M^*)^4 |T|^2~~
\label{trans}
\eeq
with relation (\ref{sig1}). 

It was found that in relativistic heavy ion collisions the phase space 
distribution is not equilibrated during much of the reaction \cite{gait01}. 
This modifies the in-medium T-matrix and thus the mean field and the 
cross section via the intermediate state Pauli operator. To account for 
this approximation, the mean field 
is described in the Local (phase space) Configuration Approximation (LCA) 
where the anisotropic phase space is taken 
into account locally in a $2$-Fermi-sphere configuration \cite{gait01}. 
In a fully consistent 
treatment one should treat the in-medium NN scattering in the same way, 
i.e. for a scattering of a nucleon to a $2$-Fermi sphere
configuration which is not available yet. To make the discussion more 
transparent here, we will discuss 
results of transport calculations where the mean field is always treated in 
the  LCA approximation. 

The connection of the microscopic BUU approach to a hydrodynamical
picture is established via macroscopic transport coefficients such 
as the shear viscosity $\eta$ and the heat capacity \cite{dani84}.  
Though defined close to the equilibrium limit, transport coefficients
give insight in the bulk properties of excited nuclear matter. 
In particular the shear viscosity is closely connected with the
in-medium cross section. Quantum effects enter via the final state
Pauli blocking for final states and for intermediate states in the BS-equation
(\ref{BS}). The cross section is not only reduced
at finite density but also becomes more isotropic which increases the 
average momentum transfer and reduces the viscosity of
the system. 

It is of great interest where such effects are experimentally
accessible in heavy ion reactions. While the close connection of 
collective flow observables to the mean field is well established 
\cite{fopi1,dani00,larionov00,gait01}, the 
experimental access to in-medium modifications of the cross section is 
not so straightforward. Stopping observables are in principle natural
tools to study the effectiveness of binary collisions. However, the
comparison to experimental rapidity distributions up to now did not 
deliver conclusive information on  in-medium modifications of the
cross section. 

Very recently, the FOPI collaboration at GSI performed a systematic 
analysis of nuclear stopping in the most central reactions over a wide
energy range form $0.1\div 1.5$ AGeV \cite{fopi5}. The complete phase
space coverage allowed not only the measurement of 
longitudinal rapidity distributions but also those transverse to 
the beam direction. The data showed that even in most
central reactions the system is not completely equilibrated which is
reflected in a broader longitudinal distribution. A quantitative
measure for the equilibration was introduced by Reisdorf et al. 
\cite{fopi5} by the ratio
\beq
var_{\rm tl} = \frac{\sigma^2(y)}{\sigma^2(z)}
\label{vartl}
\eeq
of the longitudinal $(z)$ and transverse $(y)$ variances of the rapidity 
distributions. Hence, 
a value of $var_{\rm tl}\le 1$ indicates incomplete
stopping. It was further observed that stopping (in central reactions)
and transverse flow (in semi-peripheral reactions) show a strong
correlation in that the excitation function for both observables has a
broad maximum around 0.4-0.6 AGeV \cite{fopi5}.    
\begin{figure}[t]
\unitlength1cm%
\begin{picture}(8.,8.)%
\put(2.5,0.3){\makebox{\epsfig{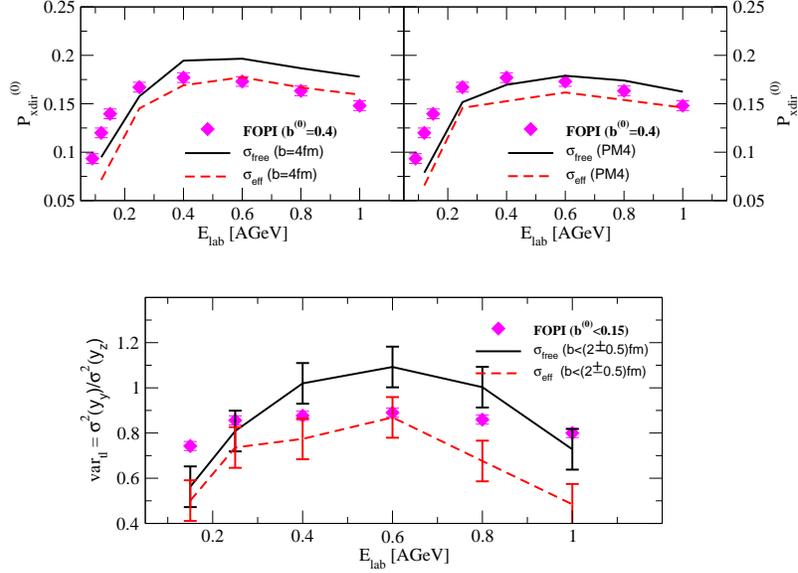}}}%
\end{picture}%
\caption{Energy dependence of the scaled directed flow %
$P_{xdir}^{(0)}=P_{xdir}/P_{\rm proj}$ (top panels) %
and the degree of stopping $var_{\rm tl}$ (bottom) for Au+Au collisions. %
The directed flow and the observable $var_{\rm tl}$ are %
experimentally determined for the scaled impact parameter %
$b^{(0)}=b/b_{max} \approx 0.4$ ($b_{max}$ being the maximal value of the impact parameter 
defined as $b_{max}=1.15(A_{\rm proj}^{1/3}+A_{\rm targ}^{1/3})$) 
corresponding to the magnitude of maximal global %
flow and for $b^{(0)} < 0.15$ in semi-central and central Au+Au 
collisions, respectively. 
Transport calculations using free (solid lines) and in-medium NN 
cross sections (dashed lines) are compared to data from the 
FOPI collaboration \protect\cite{fopi5} (filled diamonds). 
The directed flow is theoretically determined 
for a fixed impact parameter of $b=4$ fm (top-left panel) and for the centrality class 
PM4 ($b=[2,6]$ fm) (top-right panel). The observable $var_{\rm tl}$ (bottom panel) 
is theoretically calculated for $b<2.5$ fm with a variation of $\delta b=0.5$ fm 
in the impact parameter range.}
\label{fig2}
\end{figure}

Fig.~\ref{fig2} shows the incident energy dependence of the 
observable $var_{\rm tl}$ and the scaled directed flow 
$P_{xdir}^{(0)} = \left( \sum_i {\rm sign}(y_i) p_{x}^i \right)/P_{\rm proj}$, 
where the sum runs over all charged particles. 
A comparison to experimental data \cite{fopi5} requires 
an accurate determination of the centrality class in the same 
way as in the experiment. In particular, this is important for 
the observable $var_{\rm tl}$ due to its high sensitivity to the 
centrality selection. We have performed a systematic study on the 
impact parameter dependence of the flow and stopping observables. 
$P_{xdir}^{(0)}$ was theoretically determined both for the centrality interval PM4 
(top-right panel in Fig.~\ref{fig2}), corresponding to b=2-6 fm and 
for the fixed impact parameter b=4 fm (top-left panel in Fig.~\ref{fig2}), 
corresponding to the maximum magnitude of 
directed flow. The last condition was used in determining $P_{xdir}^{(0)}$ 
experimentally \cite{fopi5}. A consistent determination of 
the stopping observable $var_{\rm tl}$  and the highest centrality class 
is more difficult due to its high sensitivity 
on the centrality selection. In Ref. \cite{fopi5} the observable 
ERAT was used for the experimental determination of central 
collisions. Comparisons with IQMD simulations \cite{iqmd} provide an 
impact parameter range $b \sim 2-2.5$ fm, consistent with previous studies 
\cite{gait01}. However, because of numerical fluctuations of the ERAT 
distribution a determination of the impact parameter to better 
than $\delta b=0.25-0.5$ fm is not possible, which 
essentially affects the final result of 
$var_{\rm tl}$. The fluctuations of $var_{\rm tl}$ 
according to these $\delta b$ are indicated in 
Fig.~\ref{fig2} (bottom) as error bars. Furthermore, as for the 
experimental data \cite{fopi5},  $var_{\rm tl}$ is determined for
central reaction ($b\leq 2.5$ fm) by Gaussian fits to the 
final rapidity distributions. The theoretical calculations 
are performed for two cases, i.e. using free and in-medium 
total and differential NN cross sections (solid and dashed lines in Fig.~\ref{fig2}, 
respectively).

Generally the microscopic calculations reproduce the
qualitative behavior of the data. The stopping power as well as the
scaled directed flow have a broad maximum as a function of beam energy. 
The use of in-medium
cross sections has only little impact on the collective flow as expected. 
The flow is slightly reduced but at high energies the data are still somewhat
overestimated. This is understood from the momentum
dependence of the nucleon-nucleus optical potential. 
Within the microscopic Dirac-Brueckner 
approach \cite{DBT1} the optical potential is too repulsive at 
energies above 0.6 AGeV and does not reproduce the empirically  
observed saturation of the optical potential \cite{gait01}. 
In heavy ion collisions this leads to an overestimation 
of the collective flow, which has been cured adopting ad-hoc 
phenomenological parameterizations of the optical potential 
\cite{larionov00}. The main reason for the decrease of the flow at high energies 
is the opening of inelastic channels (pion, kaon production etc.). 
Such in-elasticities are included in the collision part of the transport
calculation but not on the level of the mean field. The theoretical
understanding of the bare NN interaction above the inelastic threshold
is still an open issue \cite{eyser03}. 
\begin{figure}[t]
\unitlength1cm
\begin{picture}(7.,6.0)
\put(2.5,0.3){\makebox{\epsfig{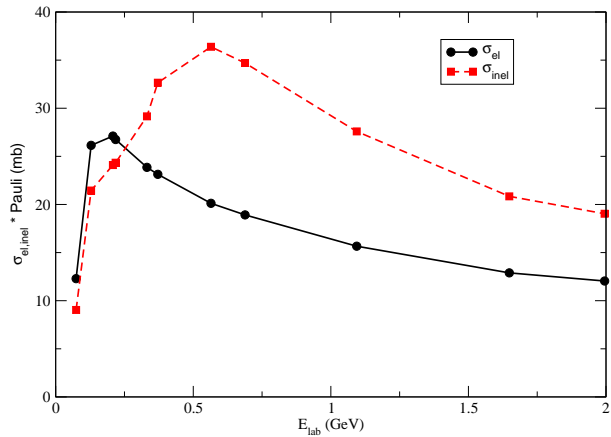}}}
\end{picture}
\caption{Elastic and inelastic NN cross section multiplied by average
Pauli blocking factors for the initial phase space configurations 
of heavy ion reactions.
}
\label{fig3}
\end{figure}

The energy dependence of $var_{\rm tl}$ can be understood more easily in a
qualitative way: at low energies Pauli blocking is most efficient, 
suppressing $2$-body collisions which leads 
to a higher transparency. With increasing energy the blocking becomes 
ineffective whereas  the 
elementary cross section drops. Hence the maximum reflects the
combination of final state Pauli blocking and a decreasing cross
section. For a quantitative description of the data the additional reduction 
of the in-medium cross section by Pauli blocking of intermediate 
states and $M^*$ scaling is seen to be essential. 

The interplay between Pauli blocking and the NN cross section can be
seen from Fig. 3. This figure shows the isospin averaged free cross 
section multiplied by average Pauli blocking factors for the initial 
local momentum configurations in the central cell of a heavy ion 
reaction. These are given by two nuclear matter currents of
density $\rho_0$ separated by the beam velocity which in 
momentum space correspond to Lorentz boosted Fermi ellipsoids. The blocking factors
are evaluated for two representative nucleons in the
centers of the ellipsoids. One clearly sees how the interplay between
Pauli blocking and decreasing cross section leads to a broad maximum in the
effective cross sections at about 0.5 GeV which appears in the same region as that of 
$var_{\rm tl}$. 

\begin{figure}[t]
\unitlength1cm
\begin{picture}(8.,7.0)
\put(2.5,0.3){\makebox{\epsfig{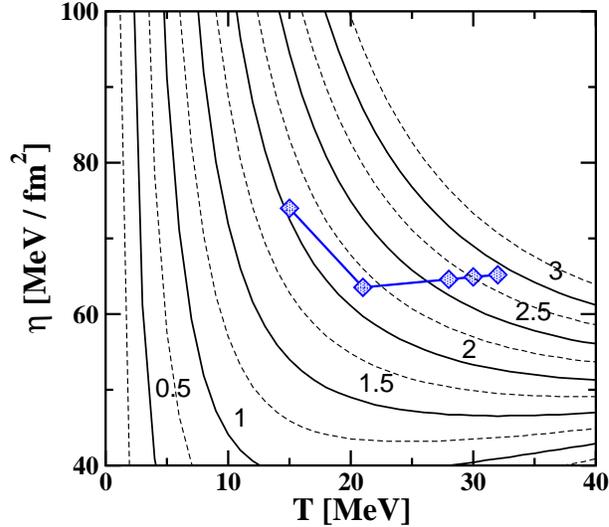}}}
\end{picture}
\caption{Shear viscosity of nuclear matter as a function of density
$\rho/\rho_0$ (indicated by labels) 
and temperature. Diamonds show the values obtained in 
central Au+Au reactions at the stage of maximal compression. The
values correspond to incident energies of 0.15, 0.25, 0.4, 0.6 and 0.8 
AGeV (with rising T).
}
\label{fig4}
\end{figure}
It is also important to obtain an understanding of the stopping
behavior in terms of nuclear matter bulk properties. However, ideal 
one-fluid hydrodynamics which is most directly related to nuclear
matter bulk properties completely fails at SIS energies \cite{fopi5}. 
In contrast, quantum effects
such as Pauli blocking of two-body collisions are essential for the
degree of stopping. In Ref. \cite{dani84} nuclear transport coefficients
have been derived from the microscopic BUU equation. In the Boltzmann
statistical limit these correspond to the first order Chapman-Enskog
coefficients. At large density the microscopic 
transport coefficients differ, however, essentially 
from the Chapman-Enskog results 
\cite{dani84}. According to \cite{dani84} the shear viscosity of a
nuclear liquid can be parameterized as a function of $\rho$ and $T$ 
\begin{equation}
\eta \sim \frac{1700}{T^2}\left(\frac{\rho}{\rho_0}\right)^2
+ \frac{22}{1 + T\cdot10^{-3}} \left(\frac{\rho}{\rho_0}\right)^{0.7} 
+ \frac{5.8\sqrt{T}}{1 + 160T^{-2}}
\label{eta}
\end{equation}
Fig. 4 shows $\eta$ according to eq. (\ref{eta}) as a function of T 
for various $\rho$ together with the values which have been extracted 
from central Au+Au
reactions. The  $T-\rho$ for which  $\eta$ is evaluated have been
determined  in the central cell at the stage of maximal
compression. The temperature was thus extracted by fits of Fermi
distribution to the local phase space as delivered from the transport
calculations \cite{temp}. From Fig. 4 one sees that $\eta$ exhibits a minimum as 
function of beam energy at $E_{\rm Lab} =0.25$ AGeV which is close to
the maximum of  $var_{\rm tl}$. 

In summary, we performed for the first time RBUU calculations where
the mean field and the in-medium cross section are consistently 
based on the same in-medium Dirac-Brueckner T-matrix. We investigated the 
recently proposed observable $var_{\rm tl}$ as a measure for the
degree of stopping and equilibration in central reactions. In-medium 
effects on the NN cross section are essential in order to reproduce
these data. Qualitatively, the behavior of the  $var_{\rm tl}$
excitation function can be understood by the interplay between the 
decreasing NN cross section and final state Pauli blocking. The 
interpretation of nuclear bulk properties in terms of transport 
coefficients indicates that the shear viscosity is minimal in the
regime where maximal stopping occurs.


\end{document}